\documentclass{amsart} 
  
\usepackage{amssymb}
\usepackage{amsmath} 
\usepackage{empheq}
\usepackage{mathrsfs} 

\copyrightinfo{2024}{Nathan Strange}

\newtheorem{theorem}{Theorem}[section]
\newtheorem{lemma}[theorem]{Lemma}
\newtheorem{proposition}[theorem]{Proposition}

\newtheorem{corollary}[theorem]{Corollary}

\theoremstyle{definition}
\newtheorem{definition}[theorem]{Definition}

\theoremstyle{remark}

\numberwithin{equation}{section}

\newcommand{\beq}[1]{\begin{equation}\label{eq:#1}}
\newcommand{\eeq}{\end{equation}}
\newcommand{\beqa}[1]{\begin{eqnarray}\label{eq:#1}}
\newcommand{\eeqa}{\end{eqnarray}}
\newcommand{\Eqref}[1]{Equation~(\ref{eq:#1})}
\renewcommand{\eqref}[1]{(\ref{eq:#1})}

\title[Analytic Solution of Navier-Stokes]{Analytic Solution of the N-Dimensional\\ Incompressible Navier-Stokes Equations}

\author[Strange]{Nathan Strange}
\email{nathan@visviva.space}
\thanks{Vis Viva Space, LLC}


\date{}

\dedicatory{}

\begin{document}

\begin{abstract}
This paper presents an analytic solution of the incompressible Navier-Stokes equations as recurrence relations for the solution's derivatives, addressing the Clay Mathematics Institute's Millennium Prize problem on Navier–Stokes existence and smoothness.
\end{abstract}

\maketitle

\section{Introduction}

This paper presents an analytic solution that addresses the Clay Math millennium problem  on the ``Existence and Smoothness of the Navier-Stokes Equation'' \cite{fefferman2000existence}. 
This analytic solution is expressed as a recurrence relation for the derivatives of the solution that may be used in a Taylor series.

The Cauchy–Kovalevskaya Theorem \cite{krantz2002primer} tells us that a system of analytic differential equations with analytic initial conditions will have an analytic solution.  With appropriate initial and boundary conditions, the Navier-Stokes equations meet these conditions.  Indeed, many of the numerical methods used in Computational Fluid Dynamics (CFD) solvers are derived in part using Taylor series approximations and, in 1986, Perry and Chong \cite{perry1986series} successfully found local analytic solutions to the Navier-Stokes equations using a computer algebra system to expand the Taylor series solutions.  Such numerical methods are very powerful, but do not provide the insight available from a general solution.

Rather than use computer algebra, the method in this paper provides arbitrary order recurrence relations for the derivatives of the solution symbolically.  These relations can be used to generate Taylor series expansions wherever the initial and boundary conditions are analytic.  This approach is extremely useful for studying the general properties of the solution and can be used to solve the Clay Math millennium problem.   

\subsection{The Clay Math Navier-Stokes Problem}

The Clay Math millennium problem as stated in ``Existence and Smoothness of the Navier-Stokes Equation'' by Fefferman \cite{fefferman2000existence} concerns the $N$-dimensional Navier-Stokes momentum differential equations for $1\le j \le N$: 
\beq{navstokes_momentum}
\frac{\partial}{\partial t} u_j(t,\vec{x}) +\sum_{k=1}^N u_k(t,\vec{x}) \frac{\partial}{\partial x_k} u_j(t,\vec{x}) =  \nu  \sum_{k=1}^N \frac{\partial^2}{\partial x_k^2}   u_k(t,\vec{x})
- \frac{\partial}{\partial x_j} p(t,\vec{x})+f_j (t,\vec{x})
\eeq
with the divergence-free constraint for incompressible flow:
\beq{navstokes_incompressible} 
\sum_{j=1}^N \frac{\partial}{\partial x_j} u_j(t,\vec{x}) = 0
\eeq
and the initial condition:
\beq{velocity_ic}
u_j(t^0,\vec{x}) = u^0_j(\vec{x})\;,\quad \vec{x} \in \mathbb{R}^N
\eeq 
Above, the $u_j(t,\vec{x}): \mathbb{R}^{N+1} \to \mathbb{R}$ are the components of the fluid velocity at $t \in \mathbb{R}$ and $\vec{x} \in \mathbb{R}^N$, $f_j (t,\vec{x}): \mathbb{R}^{N+1} \to \mathbb{R}$ is the applied specific force, and $p(t,\vec{x}): \mathbb{R}^{N+1} \to \mathbb{R}$ is the kinematic pressure (i.e. pressure divided by density).

The Clay Math problem asks for a solution that shows either the existence and smoothness of solutions or breakdown of solutions in either of two problems:\\
 
\noindent{\bf (A) Existence and smoothness of Navier–Stokes solutions on $\mathbb{R}^3$:}  

Take $\nu > 0$ and $N=3$.  Let $u^0_j(t,\vec{x})$ be any smooth, divergence free vector field satisfying:
\beq{velcond1}
\left| \frac{\partial^{\|\alpha\|}}{\partial \vec{x}^{\alpha}}u^0_j(\vec{x}) \right| 
\le \frac{C_{\alpha k}}{(1+ \|\vec{x}\|)^k} 
\eeq
where $\alpha$ is multi-index notation (see section 2.1 below) for $\alpha_1, \ldots, \alpha_N$ and $C_{\alpha k}$ is a positive constant for any $\alpha_j>0$ and $k>0$.  Take $f(t,\vec{x})$ to be identically zero.  Then there exist smooth functions $p(t,\vec{x})$ and $f(t,\vec{x})$ on $\mathbb{R}^3 \times [0,\infty)$ that satisfy \eqref{navstokes_momentum}, \eqref{navstokes_incompressible}, \eqref{velocity_ic}, and the following constraint on the flow's kinetic energy:
\beq{ke_constaint}
\int_{\mathbb{R}^3} \| \vec{u}(t,\vec{x}) \|^2 < C\;, \quad \text{for all }t\ge 0
\eeq
 
\noindent{\bf (B) Existence and smoothness of Navier–Stokes solutions in $\mathbb{R}^3/\mathbb{Z}^3$:}  

Let $u_j(t,\vec{x})$ be any smooth, divergence free vector field satisfying the following periodic intial condition for $\vec{e}_j \in \mathbb{R}^3$:
\beq{velcond2}
u^0_j(\vec{x}+\vec{e}_j) = u^0_j(\vec{x})\;, \quad p^0_j(\vec{x}+\vec{e}_j) = p^0_j(\vec{x})\;, 
\quad \text{for }1\le j \le 3
\eeq
Then there exist smooth functions $p(t,\vec{x})$ and $f(t,\vec{x})$ on $\mathbb{R}^3 \times [0,\infty)$ that satisfy \eqref{navstokes_momentum}, \eqref{navstokes_incompressible}, \eqref{velocity_ic}, and:
\beq{velcond3}
u_j(t,\vec{x}+\vec{e}_j) = u_j(t,\vec{x})\;, \quad p_j(t,\vec{x}+\vec{e}_j) = p_j(t,\vec{x})\;, 
\quad \text{for }1\le j \le 3
\eeq

In this paper I will show that (A) is true and that (B) is true if the initial conditions for velocity and pressure are analytic.

\section{Preliminaries}
 
Before attacking the Navier-Stokes problem, I need to clarify the notation that I'll be using, review some properties of real analytic functions, and describe the process for multiplying and taking partial derivatives of Taylor series.  In this section, my goal is to provide some background for readers not familiar with multivariate real analytic functions.

\subsection{Multi-Index Notation}
 
As shorthand for $N$ position coordinates, $x_1, \ldots, x_N$ I will use the vector $\vec{x}$.
When I use Greek indices such as $\alpha$, they represent ``multi-index'' notation \cite{krantz2002primer} corresponding to $\alpha_1\alpha_2\ldots\alpha_N$ indices for these $N$ coordinates.   
Below are some examples of how multi-index notation expands for $N=3$ dimensions:
\beqa{multi}
f^{\alpha} &=& f^{\alpha_1\alpha_2\alpha_3}\\
\|\alpha\|&=& \alpha_1+\alpha_2+\alpha_3\\
\frac{\partial^{\|\alpha\|}}{\partial \vec{x}^{\alpha}} f(\vec{x})&=& 
 \frac{\partial^{\alpha_1}}{\partial x_1^{\alpha_1}} \frac{\partial^{\alpha_2}}{\partial x_2^{\alpha_2}}
\frac{\partial^{\alpha_3}}{\partial x_3^{\alpha_3}} \,f(x_1,x_2,x_3) \\
\alpha! &=& \alpha_1!\alpha_2!\alpha_3!\\
(\vec{x}-\vec{x}^{\,0})^{\alpha} &=& (x_1-x^0_1)^{\alpha_1} (x_2-x^0_2)^{\alpha_2} (x_3-x^0_3)^{\alpha_3}\\
\sum_{\alpha=0}^{\infty} b^{\alpha}(\vec{x})^{\alpha} &=&  \sum_{\alpha_1=0}^{\infty} \sum_{\alpha_2=0}^{\infty} \sum_{\alpha_3=0}^{\infty} b^{\alpha_1\alpha_2\alpha_3} (x_1)^{\alpha_1} (x_2)^{\alpha_2} (x_3)^{\alpha_3}\\
\binom{n}{\alpha} &=& \frac{n!}{\alpha_1!\alpha_2!\alpha_3!}\\
\binom{\mu}{\alpha} &=& \frac{\mu_1!\mu_2!\mu_3!}{\alpha_1!\alpha_2!\alpha_3!}\\
(\vec{x})^{\alpha+\mu} &=& (x_1)^{\alpha_1+\mu_1}(x_2)^{\alpha_2+\mu_2}(x_3)^{\alpha_3+\mu_3} 
\eeqa

\subsection{Derivative Notation}  
I will use parentheses when raising a quantity to a power, e.g. $(x)^n$, and  superscripted indices to represent derivatives.  I'll use greek multi-indices for derivatives with respect to coordinates and latin indices for other derivatives.  For example:
\beq{deriv_notation}
u^{m\alpha}_j (t^0, \vec{x}^{\,0} ) = \left. \frac{\partial^m}{\partial t^m}  \frac{\partial^{\|\alpha\|}}{\partial \vec{x}^\alpha}  u_j(t,\vec{x})\, \right|_{t^0, \vec{x}^{\,0}}
\eeq
When I don't explicitly write the variables where these derivatives are evaluated, they are implied:
\beq{deriv_notation2}
u^{m\alpha}_j  \implies u^{m\alpha}_j (t^0, \vec{x}^{\,0} )
\eeq

With this notation, incrementing the superscripted index increases the order of the derivative.  I.e., if $f^n = d^n/dx^n f(x)$, then $f^{n+1} = d^{n+1}/dx^{n+1} f(x)$.  To increment one index in a set of multi-indices, I will
use $\partial^s_k(f^{\alpha})$ to increment the $k$th $\alpha_n$ by $s$, i.e.:
\beq{deriv1}
\partial^s_k \left( u^{m\alpha}_j \right) =  u^{m\alpha_1\ldots(\alpha_k+s)\ldots\alpha_N}_j 
\eeq

\subsection{Taylor Sum Notation}

Since the sum of $\frac{1}{m!} (t-t^0)^m \frac{1}{\alpha!} (\vec{x}-\vec{x}^{\,0})^{\alpha}$ terms will occur frequently in the following sections, I find it convenient to use the following shorthand:
\beq{sum1}
\Delta_{m\alpha}(t,\vec{x};t^0,\vec{x}^{\,0}) = \frac{1}{m!} (t-t^0)^m \frac{1}{\alpha!} (\vec{x}-\vec{x}^{\,0})^{\alpha}
\eeq
As in \eqref{deriv_notation2}, when the variables are omitted, they are implied:
\beq{sum2}
\Delta_{m\alpha} \implies \Delta_{m\alpha}(t,\vec{x};t^0,\vec{x}^{\,0})
\eeq
For example, the Taylor series expansions of the velocity components and pressure can be written as:
\beqa{taylor_u}
u_j(t,\vec{x}) &=&  \sum_{m = 0}^{\infty} \sum_{\alpha = 0}^{\infty}u^{m\alpha}_j(t^0, \vec{x}^{\,0} )\, \frac{1}{m!} (t-t^0)^m \frac{1}{\alpha!} (\vec{x}-\vec{x}^{\,0})^{\alpha} \label{eq:taylor_u1}\label{eq:taylor_u1} \\
& = & \sum_{m = 0}^{\infty} \sum_{\alpha = 0}^{\infty}u^{m\alpha}_j (t^0, \vec{x}^{\,0} )\, \Delta_{m\alpha}(t,\vec{x}; t^0,\vec{x}^{\,0})   \nonumber \\
& = & \sum_{m = 0}^{\infty} \sum_{\alpha = 0}^{\infty}u^{m\alpha}_j  \Delta_{m\alpha}  \nonumber \\
p(t,\vec{x}) &=&  \sum_{m = 0}^{\infty} \sum_{\alpha = 0}^{\infty}p^{m\alpha} (t^0, \vec{x}^{\,0} )\, \frac{1}{m!} (t-t^0)^m \frac{1}{\alpha!} (\vec{x}-\vec{x}^{\,0})^{\alpha} \label{eq:taylor_u2}  \label{eq:taylor_u2}\\
& = & \sum_{m = 0}^{\infty} \sum_{\alpha = 0}^{\infty}p^{m\alpha} (t^0, \vec{x}^{\,0} )\, \Delta_{m\alpha}(t,\vec{x}; t^0,\vec{x}^{\,0}) \nonumber \\
& = &\sum_{m = 0}^{\infty} \sum_{\alpha = 0}^{\infty}p^{m\alpha} \Delta_{m\alpha}\nonumber  
\eeqa

\subsection{Real Analytic Functions} 

I will use the following definition and properties of real analytic functions from Krantz and Parks \cite{krantz2002primer}.   For the propositions, I have provided outlines of alternate proofs that I hope are more accessible to readers with engineering or science backgrounds.

\begin{definition}\label{D:real_analytic}
A function $f$, with domain an open subset $X \subseteq \mathbb{R}^N$ and range $\mathbb{R}$ is called ``real analytic'' on $X$, written $f \in C^{\omega}(X)$, if for each $x_1,\ldots,x_N \in X$ the function $f$ may be represented by a convergent power series in some neighborhood of $x_1,\ldots,x_N$.
\end{definition}

 \begin{proposition}\label{P:anal_test}
 Let $f \in C^\infty(U)$ for some $U \in \mathbb{R}^N$.  The function $f$ is in fact in $C^\omega(U)$ if and only if, for each $x_1,\ldots,x_N \in U$, there is an open ball $V$, with $x_1,\ldots,x_N \in V \subseteq U$, and constants $C>0$ and $R>0$ such that the derivatives of $f$ satisfy:
 \beq{analytic_test}
 \left| f^\alpha(\vec{x}^{\,0}) \right| \le C \cdot \frac{\alpha !}{R^{\|\alpha\|}} \; ,\quad  \forall x^0_j \in V
 \eeq
 \end{proposition}
 \begin{proof} 
To derive \eqref{analytic_test}, first convert a univariate geometric series into a multivariate series using the substitution $r = \sum_{j=1}^{N} (x_j-x^0_j)/R$ and the multinomial theorem:
\beq{anal1}
\sum_{n=0}^\infty C (r)^n = \sum_{\alpha=0}^\infty \frac{C}{R^{\|\alpha\|}} \binom {\|\alpha\|}{\alpha} (\vec{x}-\vec{x}^{\,0})^{\alpha}
\eeq 
 Then via the comparison test with \eqref{anal1}, the following series converges when $ | \sum_{j=1}^{N}( x_j-x^0_j) | < R$:
\beq{anal2}
\sum_{\alpha=0}^\infty \frac{C}{R^{\|\alpha\|}}  (\vec{x}-\vec{x}^{\,0})^{\alpha}
\eeq 
The comparison test of a Taylor series: $\sum_{\alpha=0}^\infty f^\alpha(\vec{x}^{\,0})/\alpha! (\vec{x}-\vec{x}^{\,0})^{\alpha}$  with \eqref{anal2} yields \eqref{analytic_test} and the limit comparison test shows the Taylor series converges if and only if \eqref{analytic_test} is true.  
Next, to show the Taylor series converges to $f$, consider the Taylor series remainder with the constraint in \eqref{analytic_test}:
\beq{remainder}
| R_\nu(\vec{x}-\vec{x}^{\,0}) | \;=\;  
\sum_{\alpha=\nu}^\infty \left|\frac{f^\alpha(\vec{x}^{\,0})}{\alpha!} (\vec{x}-\vec{x}^{\,0})^{\alpha}\right|
 \;\le\;    
\sum_{\alpha=\nu}^\infty  \frac{C}{R^{\|\alpha\|}} \left| (\vec{x}-\vec{x}^{\,0})^{\alpha}\right|
\eeq
The right side of \eqref{remainder} is the remainder of the geometric series in \eqref{anal2} which tends to zero as $\nu \to \infty$.   Therefore if \eqref{analytic_test} is true and $f \in C^\infty(U)$, the Taylor series of $f$ converges to $f$ and $f$ is real analytic.
 \end{proof}
 
 \begin{proposition}\label{P:add_mul}
Let $U,V \subseteq \mathbb{R}^N$ be open.  If $f: U \to \mathbb{R}$ and $g: V \to \mathbb{R}$ are real analytic, then $f+g$, $f\cdot g$ are real analytic on $U \cap V$, and $f/g$ is real analytic on $U \cap V \cap \{x_j: g(x_1,\ldots,x_N) \neq 0\}$.
\end{proposition}
 \begin{proof} 
An outline of the proof: use bounds from Proposition \ref{P:anal_test} with the same $R$ for both series (i.e. an $R$ less than the radii of convergence for both series).   For addition, take the sum of two Taylor series satisfying the bounds in \eqref{analytic_test} and show that the sum also satisfies \eqref{analytic_test}.  For multiplication, use the generalized Leibniz product rule (see \eqref{mul4} below) to show the same also applies to products of Taylor series and it follows that multiplying analytic functions yields analytic functions.  As division is the inverse of multiplication, if $f(\vec{x}) = c(\vec{x}) \cdot g(\vec{x})$ is analytic and $g(\vec{x})$ is analytic and nonzero, then $c(\vec{x}) = f(\vec{x}) / g(\vec{x})$ must be analytic. 
 \end{proof}

\begin{proposition}\label{P:deriv_int}
Let $f$ be a real analytic function defined on an open subset $X \subseteq \mathbb{R}^N$.  Then $f$ is continuous and has real analytic partial derivatives of all orders.  Further, the indefinite integral of $f$ with respect to any variable is real analytic.
\end{proposition}
 \begin{proof} 
An outline of the proof: by comparison test, show that $\sum_{\alpha=0}^\infty \frac{C}{R^{\|\alpha\|}} \binom {\|\alpha\|+k}{\alpha} (\vec{x})^{\alpha}$ converges if and only if \eqref{anal1} converges.  Then via comparison test with a Taylor series, convert the condition in \eqref{analytic_test} into:
\beq{anal_test1}
 \left| f^\alpha \right| \le C \cdot \frac{(\|\alpha\|+k) !}{R^{\|\alpha\|}} \; ,\quad  \forall x_j \in V,\,  \forall k \in \mathbb{N}_0
\eeq
As differentiation of Taylor series increments indices (see 2.5 below), this is equivalent to different values of $k$ in \eqref{anal_test1} and the proposition follows for both derivatives and anti-derivatives. 
 \end{proof}

\subsection{Differentiation and Integration} If we take the derivative of each term of a Taylor series for $f(\vec{x})$ with respect to $x_k$, the $f^{\alpha}(\vec{x}^{\,0})$ is not a function of $\vec{x}$, and the derivative of $\Delta_{\alpha} (\vec{x}; \vec{x}^{\,0})$ is found by decrementing the $\alpha_k$ index.   Since we are using the 
$\alpha$ indices as dummy indices in a sum, this is equivalent to incrementing the $\alpha_k$ index on the $f^{\alpha}(\vec{x}^{\,0})$:
\beqa{diff1}
\left( \frac{\partial}{\partial x_k} \right)^s \sum_{\alpha =0}^{\infty} f^{\alpha}(\vec{x}^{\,0}) \Delta_{\alpha} (\vec{x}; \vec{x}^{\,0}) 
&=& \sum_{\alpha =0}^{\infty} f^{\alpha}(\vec{x}^{\,0}) \partial^{(-s)}_k\! \left( \Delta_{\alpha}(\vec{x}; \vec{x}^{\,0}) \right) \\
&=& \sum_{\alpha =0}^{\infty} \partial^s_k \left( f^{\alpha}(\vec{x}^{\,0})  \right) \Delta_{\alpha}(\vec{x}; \vec{x}^{\,0}) \nonumber
\eeqa
Similarly, to integrate on $x_k$, we increment the $\alpha_k$ index on $\Delta_{\alpha} (\vec{x}; \vec{x}^{\,0})$ and add $F^{\alpha}$ constants of integration:
\begin{multline} \label{eq:int2}
\int^s  \sum_{\alpha =0}^{\infty} f^{\alpha}(\vec{x}^{\,0})  \Delta_{\alpha}(\vec{x}; \vec{x}^{\,0}) (dx_k)^s = \\ \sum_{\alpha =0}^{\infty} f^{\alpha}(\vec{x}^{\,0})  \partial^s_k(\Delta_{\alpha} (\vec{x}; \vec{x}^{\,0}) ) 
 +   \sum_{\alpha_{(j\ne k)}=0}^{\infty} \, \sum_{\alpha_k =0}^{ s} F^{\alpha}(\vec{x}^{\,0}) \Delta_{\alpha}(\vec{x}; \vec{x}^{\,0})  
\end{multline}

\subsection{Multiplying Series}
To solve the Navier-Stokes equations in terms of Taylor series, we will need to multiply the infinite series.  For this, we can use the Cauchy product.   This is found by multiplying the series and collecting like terms.
\beqa{mul1}
c(t,\vec{x})&=& a(t,\vec{x}) \cdot b(t,\vec{x}) \\
&=& \left(  \sum_{p = 0}^{\infty} \sum_{\mu = 0}^{\infty} a^{p\mu} \frac{1}{p!} (t-t^0)^p \frac{1}{\mu!} (\vec{x}-\vec{x}^{\,0})^{\mu} \right)
\nonumber \\
& &   \times
\left(  \sum_{q = 0}^{\infty} \sum_{\nu = 0}^{\infty} b^{q\nu} \frac{1}{q!} (t-t^0)^q \frac{1}{\nu!} (\vec{x}-\vec{x}^{\,0})^{\nu} \right) \nonumber\\
 &=&  \sum_{p = 0}^{\infty} \sum_{\mu = 0}^{\infty}  \sum_{q = 0}^{\infty} \sum_{\nu = 0}^{\infty}  a^{p\mu}  b^{q\nu} \frac{1}{p!q!} (t-t^0)^{(p+q)} \frac{1}{\mu!\nu!}  (\vec{x}-\vec{x}^{\,0})^{(\mu+\nu)} \nonumber
 \eeqa
 Introducing new indices $m = p+q$ and $\alpha = \mu + \nu$ yields:
   \begin{multline}\label{eq:mul2}
c(t,\vec{x})= \sum_{p = 0}^{\infty} \sum_{\mu = 0}^{\infty}  \sum_{m = 0}^{\infty} \sum_{\alpha= 0}^{\infty} a^{p\mu} b^{(m-p)(\alpha-\mu)} \\
\times \frac{m!}{p!(m-p)!} \frac{\alpha!}{\mu!(\alpha-\nu)!} 
 \frac{1}{m!} 
  (t-t^0)^{m}\frac{1}{\alpha!} (\vec{x}-\vec{x}^{\,0})^{\alpha}
  \end{multline}
  This can be rewritten with multi-index binomial coefficients and the $\Delta_{m\alpha}$ notation from \eqref{sum2} as:
  \beq{mul3}
c(t,\vec{x})= \sum_{m = 0}^{\infty} \sum_{\alpha= 0}^{\infty}  \sum_{p = 0}^{m} \sum_{\mu = 0}^{\alpha}  \binom{m}{p} \binom{\alpha}{\mu} a^{p\mu} b^{(m-p)(\alpha-\mu)} 
   \Delta_{m\alpha}
  \eeq
  
  The coefficients of the Taylor expansion of $c(t,\vec{x}) = \sum_{m = 0}^{\infty} \sum_{\alpha= 0}^{\infty}   c^{m\alpha}   \Delta_{m\alpha} $ are:
  \beq{mul4}
  c^{m\alpha} = \sum_{p = 0}^{m} \sum_{\mu = 0}^{\alpha}  \binom{m}{p} \binom{\alpha}{\mu} a^{p\mu} b^{(m-p)(\alpha-\mu)}
  \eeq
The $c^{m\alpha}$ in \eqref{mul4} correspond to the generalized Leibniz product rule for the derivatives of the product of $a(t,\vec{x}) \cdot b(t,\vec{x})$.

\section{Series Solution of the Navier-Stokes Equations}

In this section, I will develop recurrence relations for derivatives of the analytic solution of the Navier-Stokes equations in 
\eqref{navstokes_momentum} and \eqref{navstokes_incompressible}.   In the next section, I will use this solution to address the Clay Math problem.  
Although the Clay Math problem may be the hook that brought you to read this far, I think the recurrence relations for the analytic solution in this section 
are the more interesting result.

I start with candidate series for real analytic functions  $u_j(t,\vec{x})$ and $p(t,\vec{x})$ from \eqref{taylor_u1} and \eqref{taylor_u2}, and with a real analytic function for applied force: $f_j (t,\vec{x})= f_j^{m\alpha} \Delta_{m\alpha}$.
I will then substitute these series expansions into \eqref{navstokes_momentum} and \eqref{navstokes_incompressible} and 
solve for the derivatives of the unknown analytic functions.

\subsection{Divergence-Free Constraint}

Substitute the Taylor series expansion of $u_j(t,\vec{x})$ from \eqref{taylor_u1}  into \eqref{navstokes_incompressible} using the derivative index increment notation in \eqref{deriv1}:
\beq{navstokes2b} 
\sum_{j=1}^N \frac{\partial}{\partial x_j} u_j(t,\vec{x}) =   \sum_{j=1}^N \partial^1_j\left( u^{m\alpha}_j \right)  \Delta_{m\alpha}
\eeq
As the $\Delta_{\alpha}$ are orthogonal functions, this yields a constraint on the derivatives of $u_j(t,\vec{x})$:
\beqa{navstokes3b}  
 \sum_{j=1}^N\partial^1_j\left( u^{m\alpha}_j \right) = 0 
\eeqa 

\subsection{Pressure Poisson Equation}
  
 Even if given a velocity initial condition, $u_j(t^0,\vec{x})$, that satisfies the divergence free condition in \eqref{navstokes_incompressible}, $u_j(t,\vec{x})$ must meet this constraint for $t>t^0$.   As an alternative to \eqref{navstokes_incompressible}, this condition can be enforced with pressure and the pressure Poisson equation.  

 To derive the pressure Poisson equation, first take sum the $\partial/\partial x_j$ derivatives of the Navier-Stokes momentum equations \eqref{navstokes_momentum} :
 \begin{multline}\label{eq:pressure_poisson1}
\sum_{j=1}^N \frac{\partial}{\partial x_j} \left( 
\frac{\partial}{\partial t} u_j(t,\vec{x}) +\sum_{k=1}^N u_k(t,\vec{x}) \frac{\partial}{\partial x_k} u_j(t,\vec{x}) 
\right) = \\
  \sum_{j=1}^N \frac{\partial}{\partial x_j} \left( 
\nu  \sum_{k=1}^N \frac{\partial^2}{\partial x_k^2}  u_k(t,\vec{x})
- \frac{\partial}{\partial x_j} p(t,\vec{x})+f_j (t,\vec{x})
\right)
\end{multline}
Assuming $u(t,\vec{x})$ and $p(t,\vec{x})$ have continuous second derivatives, their second derivatives are symmetric and their $\partial/\partial x_j$ derivatives can be reordered:
 \begin{multline} \label{eq:pressure_poisson2}
\frac{\partial}{\partial t} \sum_{j=1}^N \frac{\partial u_j}{\partial x_j}   +
\sum_{k=1}^N   \left( \sum_{j=1}^N \frac{\partial}{\partial x_j} u_k(t,\vec{x}) \frac{\partial}{\partial x_k}  u_j(t,\vec{x}) 
+ u_k(t,\vec{x})\frac{\partial}{\partial x_k}\sum_{j=1}^N \frac{\partial u_j}{\partial x_j}  \right)
\\ =
\nu \sum_{j=1}^N \frac{\partial}{\partial x_j} \sum_{k=1}^N \frac{\partial^2 u_k}{\partial x_k^2}  
-   \sum_{j=1}^N \frac{\partial^2}{\partial x_j^2}  p(t,\vec{x})+  \sum_{j=1}^N \frac{\partial}{\partial x_j}  f_j (t,\vec{x})  
\end{multline}
This assumption is justified later when I show that $u(t,\vec{x})$ and $p(t,\vec{x})$ are analytic.
Per \eqref{navstokes_incompressible}, cancel the $\sum_{j=1}^N \partial u_j /\partial x_j = 0$ and \eqref{pressure_poisson2} becomes the following pressure Poisson equation:
\beq{pressure_poisson3}
 \sum_{j=1}^N \frac{\partial^2}{\partial x_j^2} p(t,\vec{x}) = \sum_{j=1}^N \frac{\partial}{\partial x_j}  f_j (t,\vec{x}) -  \sum_{j=1}^N \sum_{k=1}^N\frac{\partial}{\partial x_j} u_k(t,\vec{x}) \frac{\partial}{\partial x_k} u_j(t,\vec{x}) 
\eeq

\Eqref{pressure_poisson3} consists only of derivative, multiplication, and addition operations, therefore, with Propositions \ref{P:add_mul} and \ref{P:deriv_int}, it may be used to show that pressure is analytic with analytic initial conditions.   It also can be used to provide a recurrence relation for pressure derivatives.

\begin{lemma}\label{L:pressure}
For the Navier-Stokes equations in \eqref{navstokes_momentum} and \eqref{navstokes_incompressible}, if the velocities $u_j(t,\vec{x}) \in C^\omega(X)$ and forces $f_j(t,\vec{x}) \in C^\omega(X)$ for $t, \vec{x} \in X \subseteq \mathbb{R}^{N+1}$, then $p(t,\vec{x}) \in C^\omega(X)$ and the derivatives of pressure, $p^{m\alpha}$, satisfy:
\begin{multline}\label{eq:pressure_poisson_derivs}  
 \sum_{j=1}^N\partial^2_j \left( p^{m\alpha} \right) = \sum_{j=1}^N \partial^1_j \left( f_j^{m\alpha} \right) \\-  \sum_{j=1}^N \sum_{k=1}^N
  \sum_{p = 0}^{m} \sum_{\mu = 0}^{\alpha} \binom{m}{p} \binom{\alpha}{\mu} 
 \partial^1_j \left(u_k^{p\mu} \right) \partial^1_k \left( u_j^{(m-p)(\alpha-\mu)} \right)  
\end{multline}
\end{lemma}
\begin{proof}
By Propositions \ref{P:add_mul} and \ref{P:deriv_int}, if $u_j(t,\vec{x}) \in C^\omega(X)$ and $f_j(t,\vec{x}) \in C^\omega(X)$ for $t, \vec{x} \in X \subseteq \mathbb{R}^{N+1}$, then the right hand side of \eqref{pressure_poisson3} is analytic and $p(t,\vec{x}) \in C^\omega(X)$.
Substituting in Taylor series for $u(t,\vec{x})$ and $p(t,\vec{x})$ into \eqref{pressure_poisson3} and applying the multiplication rule from \eqref{mul4} yields:
 \begin{multline}
 \sum_{m=1}^\infty  \sum_{\alpha =1}^\infty  \left( \sum_{j=1}^N \sum_{j=1}^N\partial^2_j \left( p^{m\alpha} \right) \right)
 \Delta_{m\alpha} =   \sum_{m=1}^\infty  \sum_{\alpha =1}^\infty \left(\sum_{j=1}^N \partial^1_j \left( f_j^{m\alpha} \right) 
 \right. \\ \left.
 -  \sum_{j=1}^N \sum_{k=1}^N
  \sum_{p = 0}^{m} \sum_{\mu = 0}^{\alpha} \binom{m}{p} \binom{\alpha}{\mu} 
 \partial^1_j \left(u_k^{p\mu} \right) \partial^1_k \left( u_j^{(m-p)(\alpha-\mu)} \right)  \right)  \Delta_{m\alpha}
\end{multline}
As the $\Delta_{m\alpha}$ are orthogonal functions, matching $\Delta_{m\alpha}$ coefficients
then yields the relation for the pressure derivatives in \eqref{pressure_poisson_derivs}.
\end{proof}

\subsection{Solving the Momentum Equations}
 
Substituting candidate Taylor series for velocity and pressure into  \eqref{navstokes_momentum} provides the analytic solution for the Navier-Stokes equations.

\begin{theorem}\label{T:momentum}
For the Navier-Stokes equations in \eqref{navstokes_momentum}, if the forces $f_j(t,\vec{x}) \in C^\omega(X)$ and initial velocities $u_j(t^0,\vec{x}) \in C^\omega(X)$ for $t, t^0, \vec{x} \in X \subseteq \mathbb{R}^{N+1}$, then $u(t,\vec{x}) \in C^\omega(X)$ and $p(t,\vec{x}) \in C^\omega(X)$. Additionally the time derivatives of velocity, $u^{m\alpha}$, satisfy:
\begin{multline}\label{eq:momentum_sol}
 u_j^{(m+1)\alpha}  
 =  \nu  \sum_{k=1}^N \partial^2_k \left( u_k^{m\alpha}  \right)  
  - \partial^1_j \left( p^{m\alpha}  \right)  +  f_j^{m\alpha} \\- 
 \sum_{p = 0}^{m} \sum_{\mu = 0}^{\alpha} \binom{m}{p} \binom{\alpha}{\mu} 
\sum_{k=1}^N  u_k^{p\mu}  \partial^1_k \left( u_j^{(m-p)(\alpha-\mu)}\right)
\end{multline}
\end{theorem}
\begin{proof}
By Propositions \ref{P:add_mul} and \ref{P:deriv_int}, if $u_j(t^0,\vec{x}), f_j(t,\vec{x}), p(t^0,\vec{x}) \in C^\omega(X)$ for $t^0, t, \vec{x} \in X \subseteq \mathbb{R}^{N+1}$, then the right hand side of \eqref{navstokes_momentum} is analytic and $u(t,\vec{x}) \in C^\omega(X)$.  By Lemma \ref{L:pressure}, $p(t^0,\vec{x}) \in C^\omega(X)$ if the force and initial velocity are analytic.  If \eqref{momentum_sol} is also true, then the time derivatives of $u(t,\vec{x})$ and $p(t,\vec{x})$ are consistent with Proposition \ref{P:anal_test} and $u(t,\vec{x}), p(t,\vec{x}) \in C^\omega(X)$.

To solve \eqref{navstokes_momentum}, I will substitute in Taylor series for  $u_j(t,\vec{x})$, $p(t,\vec{x})$, and $f_j(t,\vec{x})$, and then write it
in a form summing terms of like orders.  I.e, transform \eqref{navstokes_momentum} into this form:
\beq{nav_goal_1}
A^{m\alpha}_j \Delta_{m\alpha} +
B^{m\alpha}_j \Delta_{m\alpha}=
 C^{m\alpha}_j   \Delta_{m\alpha}
- D^{m\alpha}_j   \Delta_{m\alpha}
+ f^{m\alpha}_j \Delta_{m\alpha}
\eeq
First, find $A^{m\alpha}_j$ by substituting \eqref{taylor_u1} for $u_j(t,\vec{x})$:
\beq{navstokes2}
\frac{\partial}{\partial t} u_j(t,\vec{x})  = \sum_{m = 0}^{\infty} \sum_{\alpha = 0}^{\infty} u_j^{(m+1)\alpha} \Delta_{m\alpha} \quad \implies \quad A^{m\alpha}_j =  u_j^{(m+1)\alpha} 
\eeq
For the next term: 
\begin{multline}\label{eq:navstokes3}
\sum_{k=1}^N u_k(t,\vec{x}) \frac{\partial}{\partial x_k} u_j(t,\vec{x}) 
\\ =  \sum_{k=1}^N \left(  \sum_{p = 0}^{\infty} \sum_{\mu = 0}^{\infty}
u_k^{p\mu} \frac{1}{p!}    \Delta_{p\mu}  \right)    \left( \sum_{q = 0}^{\infty} \sum_{\nu = 0}^{\infty}  \partial^1_k \left( u_j^{q\nu}\right)\Delta_{q\nu}  \right)  
\end{multline}
Using the multiplication rule from \eqref{mul4}:
\beq{navstokes3a}
B^{m\alpha}_j =   \sum_{m = 0}^{\infty} \sum_{\alpha = 0}^{\infty} \sum_{p = 0}^{m} \sum_{\mu = 0}^{\alpha} \binom{m}{p} \binom{\alpha}{\mu} 
\sum_{k=1}^N  u_k^{p\mu}  \partial^1_k \left( u_j^{(m-p)(\alpha-\mu)}\right) 
\eeq

$C^{m\alpha}_j$ is the Laplacian of $u_k(t,\vec{x})$:
\beq{navstokes4}
\nu  \sum_{k=1}^N \frac{\partial^2}{\partial x_k^2}  u_k(t,\vec{x}) =  \sum_{m = 0}^{\infty} \sum_{\alpha = 0}^{\infty} \nu  \sum_{k=1}^N \partial^2_k \left( u_k^{m\alpha}  \right)  \Delta_{m\alpha} 
\eeq
\beq{navsotkes4a}
C^{m\alpha}_j =   \nu  \sum_{k=1}^N \partial^2_k \left( u_k^{m\alpha}  \right)    
\eeq

$D^{m\alpha}_j$ is the partial derivative of $p(t,\vec{x})$:
\beq{navstokes5}
 \frac{\partial}{\partial x_j} p(t,\vec{x}) =    \sum_{m = 0}^{\infty} \sum_{\alpha = 0}^{\infty}  \partial^1_j \left( p^{m\alpha}  \right)   \Delta_{m\alpha}
\eeq
\beq{navstokes5a}
D^{m\alpha}_j =    \sum_{m = 0}^{\infty} \sum_{\alpha = 0}^{\infty}  \partial^1_j \left( p^{m\alpha}  \right)   
\eeq

\Eqref{momentum_sol} is then found from the $A^{m\alpha}_j$,$B^{m\alpha}_j$,$C^{m\alpha}_j$, and $D^{m\alpha}_j$ in \eqref{nav_goal_1} rearranged to give a recurrence relation for 
time derivatives of $u_j(t,\vec{x})$ from lower order derivatives of $u_j(0,\vec{x})$, $f(t,\vec{x})$, and $p(t,\vec{x})$.  
\end{proof}

\Eqref{momentum_sol} provides the $u_j^{(m+1)\alpha}(t^0,\vec{x})$ derivatives from the order $m$ time and order $\alpha$ spatial derivatives of velocity, pressure, and applied force.  If given analytic functions for applied force, pressure, and initial velocity, \eqref{momentum_sol} is enough to find $u_j(t,\vec{x})$.  However, if $p(t,\vec{x})$ is not available, then
the solution in Theorem \ref{T:momentum} and \eqref{momentum_sol} is underdetermined.   At first this may seem unsatisfying, but a physically meaningful solution needs to give different results with different boundary conditions.  (E.g. a brick should have different flow characteristics than an airplane.)    

Where given the time derivatives of velocity, e.g. for steady-state flow or on no-slip boundaries, only initial conditions for velocity are needed and \eqref{momentum_sol} can be solved for the spatial derivatives of $p(t,\vec{x})$:
  \begin{multline} \label{eq:navstokes7a} 
\partial^1_j \left( p^{m\alpha}  \right)  
 =  \nu  \sum_{k=1}^N \partial^2_k \left( u_k^{m\alpha}  \right)  
  - u_j^{(m+1)\alpha}  +  f_j^{m\alpha} \\ - 
 \sum_{p = 0}^{m} \sum_{\mu = 0}^{\alpha} \binom{m}{p} \binom{\alpha}{\mu} 
\sum_{k=1}^N  u_k^{p\mu}  \partial^1_k \left( u_j^{(m-p)(\alpha-\mu)}\right)
\end{multline}
 If given $f_j(t,\vec{x})$ and the time-dependent velocity on a boundary, \eqref{navstokes7a} with the zero divergence constraint \eqref{navstokes3b} may be used to get the spatial derivatives of pressure needed in \eqref{momentum_sol}.  
For example,
the boundary velocity $u_j(t,x_1^0,x_2,x_3)$ provides $u_j^{m0\alpha_2\alpha_3}$ and  \eqref{navstokes7a} provides $p^{m1\alpha_2\alpha_3}$.   Then \eqref{momentum_sol} with $p^{m1\alpha_2\alpha_3}$ provides $u_2^{m1\alpha_2\alpha_3}$ and
$u_3^{m1\alpha_2\alpha_3}$, which in \eqref{navstokes3b}  yields  $u_1^{m1\alpha_2\alpha_3}$.  Now with $u_j^{m1\alpha_2\alpha_3}$, repeat this process to get $u_j^{m2\alpha_2\alpha_3}$ and so on.

\subsection{Total System Energy}

 Since $C^\omega(X) \subset C^\infty(X)$, when the conditions of Theorem \ref{T:momentum} are met, \eqref{momentum_sol} provides the smooth pressure and velocity solutions sought in the Clay Math problem, but for version (A) of the problem I also need to show that the
 kinetic energy of the solution is bounded with the given constraints on velocity.
 
The total system specific energy, $\mathscr{E}(t,\vec{x},\vec{x}^{\,0})$, is the sum of the internal energy of the fluid, $e(t,\vec{x},\vec{x}^{\,0})$, and the kinetic energy of the fluid:
\beq{energy1}
\mathscr{E}(t,\vec{x},\vec{x}^{\,0}) = e(t,\vec{x},\vec{x}^{\,0}) + \frac{1}{2} \int_{\vec{x}^{\,0}}^{\vec{x}}   \sum_{j=1}^N \left( u_j(t,\vec{x}) \right)^2   d\vec{x}
\eeq
We can substitute Taylor series for the $u_j(t,\vec{x})$ into \eqref{energy1}:
\begin{multline}\label{eq:energy2}
\mathscr{E}(t,\vec{x},\vec{x}^{\,0}) = e(t,\vec{x},\vec{x}^{\,0}) \\+ \frac{1}{2}   \int_{\vec{x}^{\,0}}^{\vec{x}} \left( 
\sum_{m = 0}^{\infty} \sum_{\alpha = 0}^{\infty} 
\sum_{p = 0}^{m} \sum_{\mu = 0}^{\alpha}
\binom{m}{p}\binom{\alpha}{\mu}
\sum_{j=1}^N  u_j^{p\mu}  u_j^{(m-p)(\alpha-\mu)}
 \Delta_{m\alpha}  \right)  d\vec{x}
\end{multline}
The integral in \eqref{energy2} is: 
\begin{multline}\label{eq:energy4}
 \frac{1}{2} \int_{\vec{x}^{\,0}}^{\vec{x}}   \sum_{j=1}^N \left( u_j(t,\vec{x}) \right)^2  d\vec{x} =\\
  \frac{1}{2}   \sum_{m = 0}^{\infty} \sum_{\alpha = 0}^{\infty} 
\sum_{p = 0}^{m} \sum_{\mu = 0}^{\alpha} 
\binom{m}{p}\binom{\alpha}{\mu}
\sum_{j=1}^N  u_j^{p\mu} u_j^{(m-p)(\alpha-\mu)}   \Delta_{m(\alpha+1)} 
\end{multline}

The derivatives of the specific energy are then given by:
\beq{energy5} 
\mathscr{E}^{m(\alpha+1)} = e^{m(\alpha+1)} 
+   \frac{1}{2}  
\sum_{p = 0}^{m} \sum_{\mu = 0}^{\alpha} 
\binom{m}{p}\binom{\alpha}{\mu}
\sum_{j=1}^N  u_j^{p\mu} u_j^{(m-p)(\alpha-\mu)}   
\eeq

By Propositions \ref{P:add_mul} and \ref{P:deriv_int} we know that \eqref{energy5} is consistent with an analytic total energy function so long as the internal energy and velocity are also analytic.


\subsubsection{Bounds on Kinetic Energy}

As the kinetic energy integral in \eqref{energy4} is analytic with a convergent Taylor series, it is bounded when $t, \vec{x}$ are finite, but version (A) of the Clay Math problem seeks bounded energy with unbounded $t$ and $\vec{x}$, but with bounded velocity derivatives $u_j(t,\vec{x})$.  In this section I will show for bounded velocity derivatives, kinetic energy's derivatives  are also bounded.

Let's start by looking at the kinetic energy derivatives from \eqref{energy5}:
\beq{ke2}
T^{m(\alpha+1)} =  \frac{1}{2}  
\sum_{p = 0}^{m} \sum_{\mu = 0}^{\alpha} 
\binom{m}{p}\binom{\alpha}{\mu}
\sum_{j=1}^N  u_j^{p\mu} u_j^{(m-p)(\alpha-\mu)}  
\eeq
Next, where the $u_j(t,\vec{x})$ are analytic, they are bound by Proposition \ref{P:anal_test}:
\beq{u_bound}
\left| u^{m\alpha}_j \right| \le \frac{U}{(R)^{(m+\|\alpha\|)}} m! \alpha!
\eeq
Using this bound in the right side of \eqref{ke2}:
 \beq{ke3}
\frac{1}{2}  
\sum_{p = 0}^{m} \sum_{\mu = 0}^{\alpha} 
\binom{m}{p}\binom{\alpha}{\mu} N
 \frac{U}{(R)^{(p+\|\mu\|)}} p! \mu! \frac{U}{(R)^{(m+\|\alpha\|-p-\|\mu\|)}} (m-p)! (\alpha-\mu)! 
\eeq
which simplifies to:
 \beq{ke4}
 \left| T^{m(\alpha+1)} \right| \le 
 \frac{N(U)^2}{2(R)^{(m+\|\alpha\|)}} (m+1)!  (\alpha+1)! 
\eeq
We can change th4 $\alpha+1$ index to $\alpha$ on boht sides, and, as in \eqref{anal_test1}, the comparison with $(m+k)!$ is the same for an $k$, so this simplifes to:
 \beq{ke5}
 \left| T^{m\alpha} \right| \le 
 \frac{N(U)^2}{2(R)^{(m+\|\alpha\|)}} m!  \alpha!  \le  \frac{C}{(R)^{(m+\|\alpha\|)}} m!  \alpha! 
\eeq
Where $C \ge N(U)^2/2$ and Proposition \ref{P:anal_test} also holds for the kinetic energy.
We also could have deduced this result from the Cauchy–Kovalevskaya Theorem, that an analytic differential equation with analytic initial conditions will have
 an analytic solution.  But now we also have a relation between the bounds for velocity and kinetic energy.

\section{Clay Math Existence and Smoothness Problem}

I now have what is needed to prove version (A) of the Clay Math problem:
\begin{corollary}
Take $\nu > 0$ and $N=3$.  Let $u_j(t,\vec{x})$ be any smooth, divergence free vector field satisfying:
\beq{velcond1a}
\left| \frac{\partial^{\|\alpha\|}}{\partial \vec{x}^{\alpha}}u^0_j(\vec{x}) \right| 
\le \frac{C_{\alpha k}}{(1+ \|\vec{x}\|)^k} 
\eeq
where $C_{\alpha k}$ is a positive constant for any $\alpha_j$ and $k>0$.  Take $f(t,\vec{x})$ to be identically zero.  Then there exist smooth functions $p(t,\vec{x})$ and $f(t,\vec{x})$ on $\mathbb{R}^3 \times [0,\infty)$ that satisfy \eqref{navstokes_momentum}, \eqref{navstokes_incompressible}, \eqref{velocity_ic} and the following constraint on the flow's kinetic energy for some constant $C \ge 0$:
\beq{ke_constaint2}
\int_{\mathbb{R}^3} \| \vec{u}(t,\vec{x}) \|^2 d\vec{x} \;\le\;C \;, \quad \forall t\ge 0
\eeq
\end{corollary}
\begin{proof}
The initial velocity condition in \eqref{velcond1a} is consistent with Proposition \ref{P:anal_test}:
\beq{velcond1b}
\left| \frac{\partial^{\|\alpha\|}}{\partial \vec{x}^{\alpha}}u^0_j(\vec{x}^{\,0}) \right| 
\;\le\; \frac{C_{\alpha k}}{(1+ \|\vec{x}^{\,0}\|)^k}  
\;\le\; \frac{U}{(R)^{\|\alpha\|}} \alpha!
\eeq
Therefore the initial condition $u_j(t^0,\vec{x})=u^0_j(\vec{x})$ is analytic for any $\vec{x} \in \mathbb{R}^N$.  
As force is identically zero, by Theorem \ref{T:momentum}, $u_j(t,\vec{x}), p(t,\vec{x}) \in C^\omega(\mathbb{R}^N) \subset C^\infty(\mathbb{R}^N)$.

To show that kinetic energy is bounded, first expand $u(t,\vec{x})$ as a univariate Taylor series in $t$ alone:
\beq{alt_taylor}
u_j(t,\vec{x}) = \sum_{m=0}^\infty u_j^m (t^0,\vec{x}) \Delta_m(t;t^0), \quad \text{where: }u_j^m (t^0,\vec{x}) = \left. \left(\frac{\partial}{\partial t}\right)^m u_j(t,\vec{x}) \right|_{t=t^0}
\eeq
The integral in \eqref{ke_constaint2} can be written with this series as:
\beq{ke_constaint}
\int_{\mathbb{R}^3} \| \vec{u}(t,\vec{x}) \|^2 d\vec{x}  \;=\; 
\sum_{m=0}^\infty \left(
\int_{\mathbb{R}^3} \| \vec{u}^m(t^0,\vec{x}) \|^2 d\vec{x} \right)
\Delta_m(t;t^0) \;=\; T^m(t^0) \Delta_m 
\eeq

Next, let $C_k = \limsup_{\alpha \to \infty} C_{k\alpha}$ and assume the bounds on the $u^{0\alpha}$ derivatives also hold for the $u^{m\alpha}$ derivatives:
\beq{goal}
\left| u_j^m (t^0,\vec{x}) \right| \le  \frac{C_{k}}{(1+ \|\vec{x}\|)^k} 
\eeq
Then \eqref{velcond1a} with \eqref{pressure_poisson_derivs} from Lemma \ref{L:pressure}, yields an upper bound on the pressure derivatives: 
\beq{pressure_bound}
\left| p^{m\alpha} \right| \le \sum_{j=1}^N  \sum_{p=0}^m \frac{m!}{p!(m-p)!} \left( \frac{C_{k}}{(1+ \|\vec{x}\|)^k}   \right)^{p} 
\left( \frac{C_{k}}{(1+ \|\vec{x}\|)^k}   \right)^{(m-p)} 
\le N \left( \frac{C_{k}}{(1+ \|\vec{x}\|)^k}   \right)^m
\eeq
Using \eqref{goal} and \eqref{pressure_bound} with \eqref{momentum_sol} from Theorem \ref{T:momentum} provides:
\begin{multline}
\left| u_j^m (t^0,\vec{x}) \right| \le
\nu  \sum_{k=1}^N  \frac{C_{k}}{(1+ \|\vec{x}\|)^k}  
 + N \left( \frac{C_{k}}{(1+ \|\vec{x}\|)^k}   \right)^m\\ +
 \sum_{k=1}^N \sum_{p = 0}^{m} \frac{m!}{p!(m-p)!} 
\left( \frac{C_{k}}{(1+ \|\vec{x}\|)^k}   \right)^{p} 
\left( \frac{C_{k}}{(1+ \|\vec{x}\|)^k}   \right)^{(m-p)}
\end{multline}
This is consistent with the assumption in \eqref{goal}:
\beq{goal_met}
\left| u_j^m (t^0,\vec{x}) \right| \;\le\; \nu N \frac{C_{k}}{(1+ \|\vec{x}\|)^k}  + 2N \left( \frac{C_{k}}{(1+ \|\vec{x}\|)^k}   \right)^m
\;\le\; \frac{C^\prime_{\alpha k}}{(1+ \|\vec{x}\|)^k} 
\eeq
Putting this constraint into \eqref{ke_constaint} yields:
\beq{kebound}
T^m(t^0) \le N \int_{\mathbb{R}^3} \left( \frac{C^\prime_{k}}{(1+ \|\vec{x}\|)^k}  \right)^2 d\vec{x}
\eeq
However this bound means the integral in $T^m(t^0)$ evaluates to a constant, as in this improper integral:
\begin{multline}\label{eq:ybound}
 \int_{-\infty}^{\infty} \left(\frac{1}{(1+|y|)^k} \right)^2 dy =
  \lim_{a\to-\infty} \int_a^0 \frac{1}{(1-y)^{2k}} dy 
  +\lim_{b\to\infty}\int_0^b \frac{1}{(1+y)^{2k}} dy  \\
= \left(\frac{1}{2k-1} - \lim_{a\to-\infty} \frac{(1-a)^{(1-2k)}}{2k-1} \right)+
\left(  \lim_{b\to\infty} \frac{(1+b)^{(1-2k)}}{1-2k} - \frac{1}{1-2k}  \right)  
= \frac{2}{2k-1}
\end{multline}
Therefore, $T^m(t^0) \le C$ where $C \ge 0$ for all $m$ if $k\ge1$, and the flow's kinetic energy in \eqref{ke_constaint} is bounded.
\end{proof}

Theorem \ref{T:momentum} can be also used to prove version (B) of the Clay Math problem, but only for analytic initial conditions:

\begin{corollary} Let $u_j(t,\vec{x})$ be any smooth, divergence free vector field satisfying:
\beq{velcond2a}
u^0_j(\vec{x}+\vec{e}_j) = u^0_j(\vec{x})\;, \quad p^0_j(\vec{x}+\vec{e}_j) = p^0_j(\vec{x})\;, 
\quad \text{for }1\le j \le 3
\eeq
for $u^0_j(\vec{x}),\, p^0_j(\vec{x}+\vec{e}_j) \in C^{\omega}(X)$ and $t, \vec{x}, \vec{e}_j  \in X \subseteq \mathbb{R}^{N+1}$.
Then there exist smooth functions $u(t,\vec{x})$ and $p(t,\vec{x})$ that satisfy \eqref{navstokes_momentum}, \eqref{navstokes_incompressible},  and:
\beq{velcond3a}
u_j(t,\vec{x}+\vec{e}_j) = u_j(t,\vec{x})\;, \quad p_j(t,\vec{x}+\vec{e}_j) = p_j(t,\vec{x})\;, 
\quad \text{for }1\le j \le 3
\eeq
\end{corollary}
\begin{proof}
By Theorem \ref{T:momentum}, $u_j(t,\vec{x}), p(t,\vec{x}) \in C^\omega(X) \subset C^\infty(X)$.  
We set $\vec{x}^{\,1} = \vec{x}^{\,0}\vec{e}_j$.  In \eqref{momentum_sol}, if $u_j^{m\alpha}(t^0,\vec{x}^{\,1}) = u_j^{m\alpha}(t^0,\vec{x}^{\,0})$ and $p^{m\alpha}(t^0,\vec{x}^{\,1}) = p^{m\alpha}(t^0,\vec{x}^{\,0})$, the resulting Taylor series are equivalent.
\end{proof}

\section{Numerical Applications of This Solution}  

The solution in Theorem \ref{T:momentum} can be used to generate Taylor series solutions, but the convergence of these solutions is limited by the radius of convergence of these series (e.g. the upper bound on $R$ in the Proposition \ref{P:anal_test} derivative test).  Although velocity and pressure derivatives taken before reaching this limit could then be used with Theorem \ref{T:momentum} to build an analytic continuation of the solution, this approach may be more computationally intensive than existing numerical techniques for solving the Navier-Stokes equations.  
However, even when a numerical technique is more computationally efficient, it can have convergence issues that  Proposition \ref{P:anal_test} and Theorem \ref{T:momentum} can help resolve.

From Proposition \ref{P:anal_test}, a multivariate Taylor series for velocity or pressure will converge when $| t + \sum_j^N x_j | < R$.   So as time increases, the $|\sum_j^N x_j|$ that reaches the radius of convergence decreases.  This may be the cause of ``blow up time'' issues in numerical solvers mentioned by Fefferman \cite{fefferman2000existence}.

In addition, the recurrence relation \eqref{momentum_sol} from Theorem \ref{T:momentum} is underdetermined without the sufficient boundary conditions, and can also be overdetermined with incorrect boundary conditions.  When using an iterative CFD solver, such underdetermined or overdetermined boundary conditions would lead to numerical instabilities.   In these cases,  the recurrence relations in \eqref{navstokes3b},  \eqref{pressure_poisson_derivs}, and  \eqref{momentum_sol} could be used to detect and resolve such conflicts.

\section{Conclusion}

The series algebra techniques that I used here to find the analytic solution of Navier-Stokes equation can be used to solve a wide 
variety of ordinary and partial differential equations, especially when combined with Faà di Bruno's formula \cite{krantz2002primer} for the generalized chain rule.  
I have previously done this with the three-body problem \cite{strange2018analytic} and the motion of a particle in an arbitrary potential field \cite{strange2019series}.
Even 
when such analytic solutions are less computationally efficient than existing numerical methods, then can provide valuable insights into the properties of the general solution.

\bibliography{nstoked}
\bibliographystyle{amsplain}

 \end{document}